\documentclass{article}
\usepackage[dvips]{graphicx}
\begin{document}
\title{Stick-Slip Motion and Phase Transition in a Block-Spring System}
\author{Hidetsugu Sakaguchi\\
Department of Applied Science for Electronics and Materials,\\ Interdisciplinary Graduate School of Engineering Sciences,\\
 Kyushu University, Kasuga, Fukuoka 816-8580, Japan}
\maketitle
\begin{abstract}
We study numerically stick slip motions in a model of blocks and springs being pulled slowly. 
The sliding friction is assumed to change dynamically with a state variable.
The transition from steady sliding to stick-slip is subcritical in a single block and spring system. However,  we find that the transition is continuous in a long chain of blocks and springs.  The size distribution of stick-slip motions exhibits a power law at the critical point.
\end{abstract}
\section{Introduction}
Stick-slip behaviors in frictional motion have been observed in wide variety of systems such as  atomically thin lubricant films \cite{rf:1}, granular materials \cite{rf:2} and earthquake faults \cite{rf:3}.
Batista and Carlson proposed a phenomenological model of one block and one spring, in which
the transition from steady sliding to stick slip occurs via a subcritical Hopf bifurcation.\cite{rf:4}  The block is subject to a friction force, which depends on the state of the sliding surface and the velocity of the block.
A more complicated equation was proposed for the transition by Hayakawa.\cite{rf:5} 
A chain of blocks and springs was studied numerically and experimentally by Burridge and Knopoff for the earthquake faults.\cite{rf:6} Each block is connected to its neighbors with a harmonic spring and the blocks are further pulled individually forward through elastic coupling. 
Each block  is subject to friction force, which depends only on the velocity of the block.    
Carlson and Langer studied numerically a homogeneous chain of blocks and springs being pulled slowly, more in detail.  In their model, the sliding friction decreases with velocity (the velocity weakening friction) and the steady sliding motion is always unstable for all Fourier modes.\cite{rf:7,rf:8} All sizes of stick-slip events are observed and the size distribution exhibits a power law in small scales, but, great events occur more frequently than is expected from the power law. 
  
In this paper, we study a modified version of the single block and spring system of Batista and Carlson more in detail.  The friction force depends on the velocity and state variable, and the friction force increases with the velocity.   Then, 
 we study numerically a chain of blocks and springs, each of which obeys the modified model of Batista and Carlson.  
In contrast to the Carlson-Langer model, in which all blocks are pulled uniformly,  only the first block is forced to move with velocity $V$ and the other blocks are coupled only with the neighbors by harmonic springs. 
This type of forcing was treated also in the original paper by Burridge and Knopoff.\cite{rf:6}.
Elmer studied this type of train model with velocity strengthening friction and found a size distribution with complex scaling exponents.\cite{rf:9}  
Using this block-spring system, 
we find that a dynamical phase transition occurs from the steady sliding to stick slips, and 
the size distribution of the stick-slip events exhibits a power law in all scales at the critical point, however, great events occur more frequently  as the parameter is decreased from the critical point.

\section{Time Evolution and the Period of a Single Block and Spring System}
The model equation of a single block and spring system is written as
\begin{eqnarray}
M\frac{d^2x}{dt^2}&=&K(Vt-x)-F(\theta,v),\nonumber\\
\frac{d\theta}{dt}&=&\theta(1-\theta)-\alpha \theta v,
\end{eqnarray} 
 where $M,\;x$ is the mass and the position of the block, $v=dx/dt$ is the velocity, $K$ is the spring constant, and $V$ is the pulling velocity.  The state of the sliding surface is expressed by $\theta$. The value $\theta=1$ represents that the surface is in a solid state, and lower values of $\theta$ imply that the \begin{figure}[htb]
\begin{center}
\includegraphics[width=7cm]{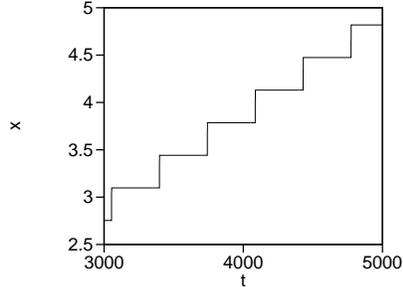}
\caption{Time evolution of $x$ in a single block and spring system eq.~(1) at $M=1,\,\alpha=1,\,\beta=1,\,\sigma=1,\,K=10,\,V=0.001$ and $F_s=3$.
} 
\label{fig:1} 
\end{center}
\end{figure} 
\begin{figure}[htb]
\begin{center}
\includegraphics[width=11cm]{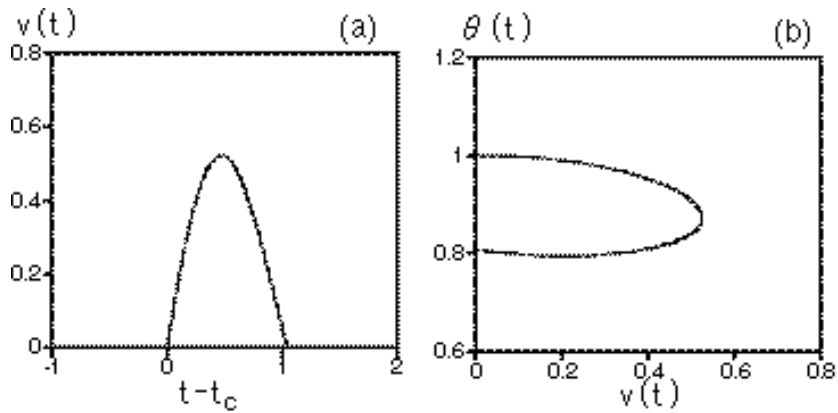}
\caption{(a) Time evolution of $v=dx/dt$ in a slip process for a single block and spring system at $M=1,\,\alpha=1,\,\beta=1,\,\sigma=1,\,K=10,\,V=0.001$ and $F_s=3$.(solid curve) and the time evolution by eq.~(3) (dashed curve). The two curves  are overlapped and the difference is invisible in this plot.  (b) Phase portraits of $(v(t),\theta(t))$ by eq.~(1) (solid curve) and by eq.~(3) (dashed curve). 
Two curves are also well overlapped. 
} 
\label{fig:2} 
\end{center}
\end{figure} 
\begin{figure}[htb]
\begin{center}
\includegraphics[width=7cm]{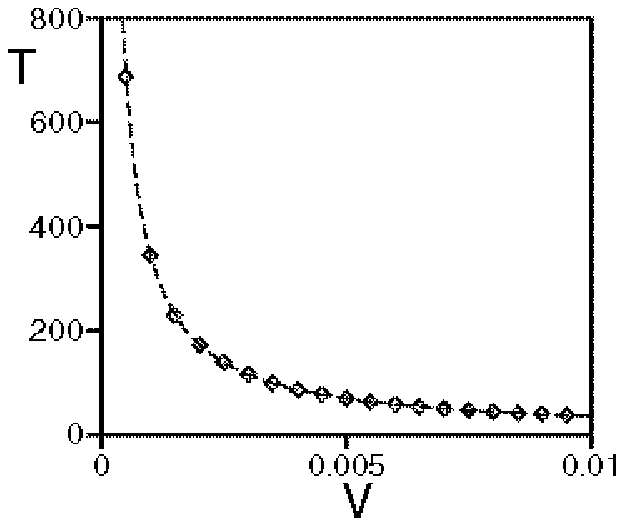}
\caption{Period of the stick slip oscillation as a function of the pulling velocity $V$. The marks denote numerical results and the dashed curve is $T=1.035+3.432/(KV)$.
} 
\label{fig:3} 
\end{center}
\end{figure} 
surface  is partially melted. The term $\alpha \theta v>0$ implies that the value of $\theta$  is decreased, i.e., the surface state becomes more melted, as the block velocity $v$ is increased.  The sliding friction force $F(\theta,v)$ is assumed to be
\[F(\theta,v)=\sigma\theta+\beta v.\]
The sliding friction contains a standard term $\beta v$ which increases in proportion to the velocity, and the 
  term $\sigma\theta$ which depends on the state variable.
There are many parameters, however, eq.~(1) can be rewritten by  scale changes: $x=\sigma/M\tilde{x},\,v=\sigma/M\tilde{v},\;V=\sigma/M \tilde{V},\,\tilde{K}=K/M,\,\tilde{\beta}=\beta/M$ and $\tilde{\alpha}=\alpha\sigma/M$ as 
\[\frac{d^2\tilde{x}}{dt^2}=\tilde{K}(\tilde{V}t-\tilde{x})-\theta-\tilde{\beta}\tilde{v},\]
\[\frac{d\theta}{dt}=\theta(1-\theta)-\tilde{\alpha}\theta \tilde{v}.\]
  That is,  the parameters $M$ and $\sigma$ can be fixed to be $M=\sigma=1$ in eq.~(1), and only
the parameters $\alpha$ and $\beta$ are important parameters. We have further fixed as $\alpha=1$ and studied the model system by changing $\beta$ in this paper for the sake of simplicity. 
We have tried several other parameter values of $\alpha$, and observed similar results.  However, the detailed parameter survey is left to the future study. 
When the block velocity is decreased to zero, the block stops. 
The block does not move until the spring force $K(Vt-x)$ goes over a limit value $F_s$ of the static friction.  We have assumed that $F_s$ is constant and it is definitely larger than $\sigma\theta$ in contrast to the model of Batista and Carlson.

A steady sliding solution is expressed as $v=v_0=V,\, \theta=\theta_0=1-V$ and $x=x_0=Vt-\{1-(1-\beta)V\}/K$. The stability can  be investigated from the linearized equation of eq.~(1)
\begin{eqnarray}
\frac{d^2\delta x}{dt^2}&=&-K\delta x-\delta\theta-\beta\delta v,\nonumber\\
\frac{d\delta\theta}{dt}&=&(1-2\theta_0-v_0)\delta\theta-\theta_0\delta v.
\end{eqnarray} 
For example, the Hopf bifurcation occurs at $V_c=0.106$ for $K=10$ and $\beta=0.08$.
The steady sliding solution becomes unstable  and only the stick-slip motion appears for $V<V_c$. However, the bifurcation is subcritical and the stick-slip motion appears in wide range of parameters. For example, the steady sliding solution is linearly stable for  any $V>0$ for $K=10$ and $\beta=1$. Stick-slip motion appears at $V=0.001$ as shown in Fig.~1. The numerical simulation was performed with the Heun method with time step 0.005. The initial condition is $x(0)=0,\,v(0)=0$ and $\theta(0)=0.1$. 
The inverse transition from stick slip to steady sliding occurs at $V=0.345$ for $\beta=1$ and $F_s=3$, as $V$ is slowly increased.
The time evolution of $v=dx/dt$ and $\theta(t)$  in the slip process hardly depends on the pulling velocity $V$, when $V$ is sufficiently small, which was pointed out in a granular experiment by Nasuno et al. \cite{rf:2}.  The time evolution in the slip process in the limit of  $V=0$ is expected to obey
\begin{eqnarray}
\frac{d^2x^{\prime}}{dt^2}&=&F_s-Kx^{\prime}-\theta-\beta v,\nonumber\\
\frac{d\theta}{dt}&=&\theta(1-\theta)-\theta v,\;\;\;\;\;\;\;\; {\rm for}\;\;t_c\,<t<\,t_e
\end{eqnarray}   
\begin{figure}[htb]
\begin{center}
\includegraphics[width=7cm]{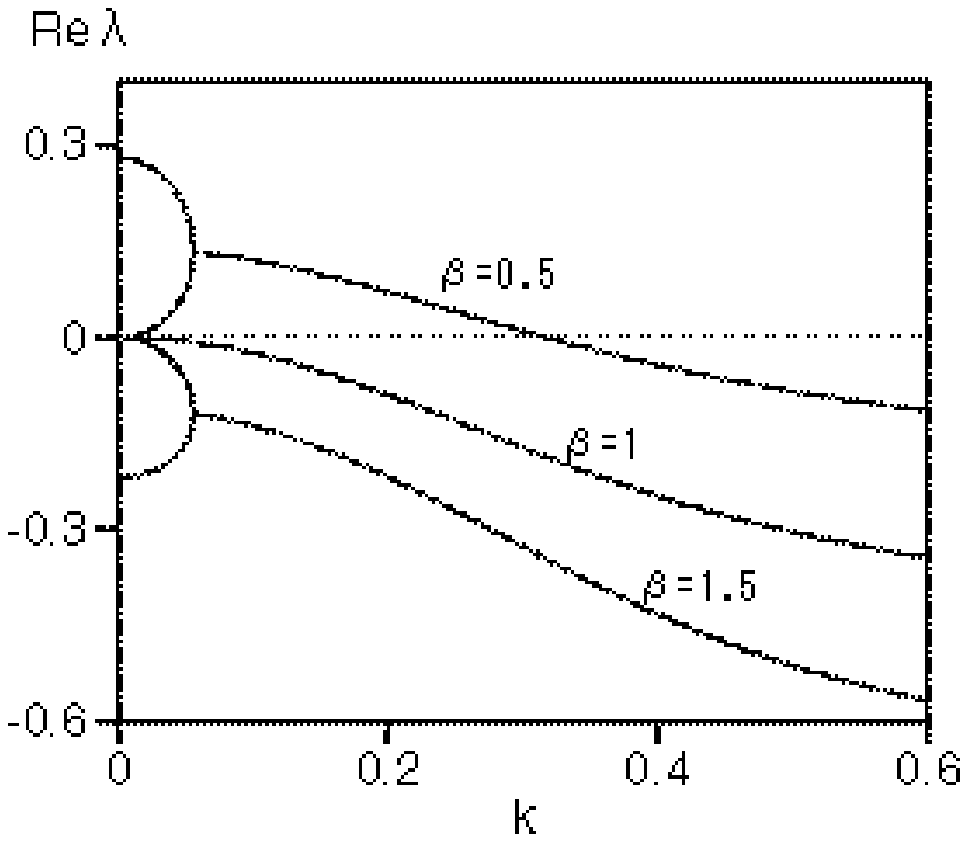}
\caption{Real parts of the largest two eigenvalues for the Fourier modes with wavenumber $k$ for eq.~(4) at $\beta=1.5,\,1$ and 0.5.
} 
\label{fig:4} 
\end{center}
\end{figure} 
\begin{figure}[htb]
\begin{center}
\includegraphics[width=11cm]{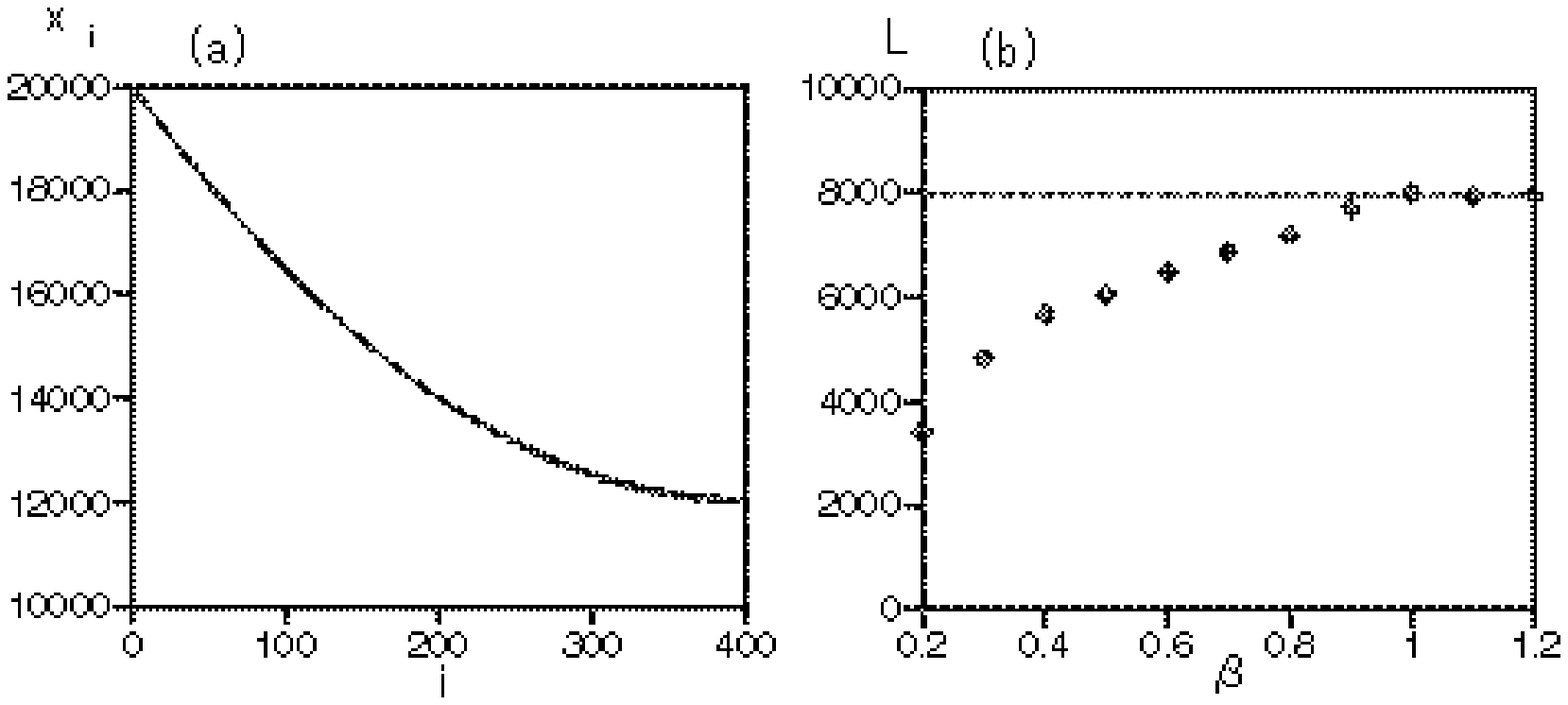}
\caption{(a) Snapshot pattern of $x_i$ at $t=2\cdot 10^7$ (solid curve) and the  steady sliding solution (dashed curve) for $\beta=1$. (b) Average total length $L=\langle x_1-x_N\rangle$ as a function of $\beta$. The dashed line denotes the total length for the steady sliding solution. 
} 
\label{fig:5} 
\end{center}
\end{figure} 
where $t_c$ and $t_e$ are the starting and ending time of a slip process, the position variable is changed as $x^{\prime}(t)=x-x(t_c)$, and the pulling velocity $V$ is neglected, and the fact that the spring force 
$K(Vt-x)$ is equal to $F_s$ at the starting time $t=t_c$ is used. The initial condition for eq.~(3) is $x^{\prime}(t_c)=0$, $v(t_c)=dx^{\prime}/dt=dx/dt=0$ and $\theta(t_c)=1$, since the state variable  is recovered to 1 during a long stick interval. 
Figure 2(a)  compares two time evolutions of $v(t-t_c)$ for eq.~(1) at $V=0.001$ and eq.~(3) for $F_s=3$, and Fig.~2(b) compares two phase portraits of $(v(t-t_c),\theta(t-t_c))$ for eq.~(1) at $V=0.001$ and eq.~(3) for $F_s=3$. 
The difference cannot be seen in these plots.  The interval of the stick depends strongly on $V$. According to the numerical simulation of eq.~(3), the time when the block stop again is $t_e=t_c+1.035$, and $F_s-Kx^{\prime}=-0.432$ at the time $t=t_e$.
The spring force $f$ is slowly increased as $f=K(Vt-x)$ in the stick interval, until the limit $f=F_s=3$ of the static friction.
The interval of the stick process is therefore evaluated as $\Delta t=3.432/(KV)$. 
The total period $T$ is therefore evaluated as $T=t_e-t_c+\Delta t=1.035+3.432/(KV)$. 
Figure 3 displays numerically obtained values of the periods of the stick slip motions and the theoretical curve.  These numerical results support the model 
equation eq.~(3) for $V=0$.  
We have shown some numerical results for specific 
parameter values, however, the qualitative  properties do not depend 
on the parameter values.

\section{Dynamic Phase Transition in a Long Chain of Blocks and Springs}
Next we consider a chain of N blocks and N springs. The model equation is written as 
\begin{eqnarray}
\frac{d^2x_i}{dt^2}&=&K(x_{i+1}-2x_i+x_{i-1})-F(\theta_i,v_i),\nonumber\\
\frac{d\theta_i}{dt}&=&\theta_i(1-\theta_i)-\theta_i v_i+D(\theta_{i+1}-2\theta_i+\theta_{i-1}),\;i=1,2,\cdots N
\end{eqnarray} 
where $x_i,\,v_i$ and $\theta_i$ represent respectively position, velocity and state variable for the $i$th block,  $K$ is the spring constant, and  a diffusion type coupling with strength $D$ is assumed in the equation of the state variable.  The sliding friction is assumed to be $F(\theta_i,v_i)=\theta_i+\beta v_i$ for nonzero $v_i$. Once the $i$th block stops,  the block does not move until the spring force  goes over the maximum value $F_s$ of the static friction as the case of the single block-spring system.   The first block is assumed to be pulled with a constant velocity as $x_0=Vt$. 
The steady sliding solution is written as $v_{i}=v_0=V, \theta_{i}=\theta_0=1-V$ and $x_{i}=Vt-\{1-(1-\beta)V\}/K\cdot \{N^2-(i-N)^2\}/2$. 
The steadily sliding state is nonuniformly strained.  
The stability of the steadily sliding solution can  be investigated from the linearized equation of eq.~(4)
\begin{eqnarray}
\frac{d^2\delta x_i}{dt^2}&=&K(\delta x_{i+1}-2\delta x_i+\delta x_{i-1})-\delta\theta_i-\beta\delta v_i,\nonumber\\
\frac{d\delta\theta_i}{dt}&=&(1-2\theta_0-v_0)\delta\theta_i-\theta_0\delta v_i+D(\delta\theta_{i+1}-2\delta\theta_{i}+\delta\theta_{i-1}).
\end{eqnarray}  
The real parts of the largest two eigenvalues for the Fourier modes with wavenumber $k$ are shown at three parameter values 0.5,1 and 1.5 of $\beta$ for  $K=10$ and $D=5$ in Fig.~4. The steady sliding solution becomes unstable against  long wavelength perturbations for $\beta\le 1$. The parameter $\beta=1$ is a critical value.
\begin{figure}[htb]
\begin{center}
\includegraphics[width=10cm]{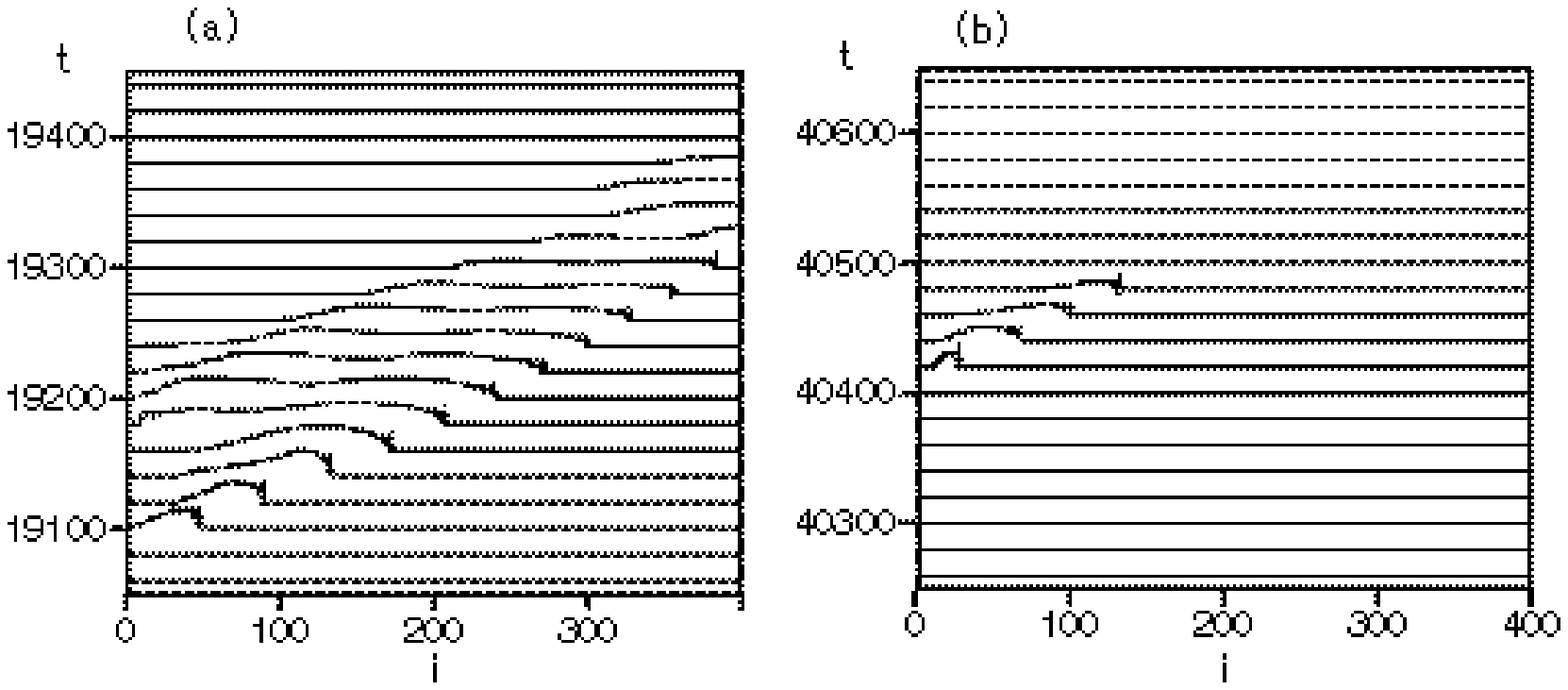}
\caption{(a)Time evolution of the displacement $\Delta x_i=x_i(t+\Delta t)-x_i(t)$ in a time interval $\Delta t=20$ in a large slip event for $\beta=1$.
(b)Time evolution of the displacement $\Delta x_i=x_i(t+\Delta t)-x_i(t)$ in a time interval $\Delta t=20$ in an intermediate slip event for $\beta=1$.} 
\label{fig:6} 
\end{center}
\end{figure} 
\begin{figure}
\begin{center}
\includegraphics[width=11cm]{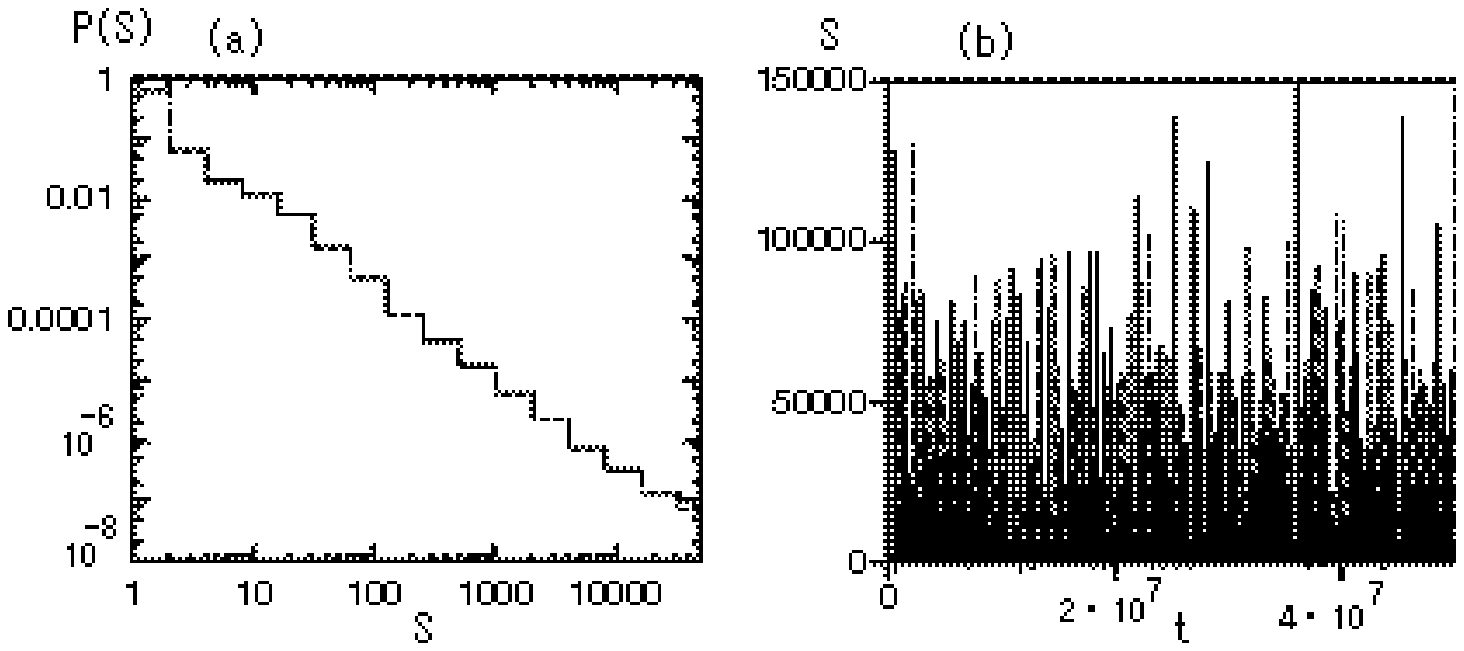}
\caption{(a) Histogram of size distribution of the total displacement $S=\sum_{i=1}^N\Delta x_i$ during each slip event at $\beta=1$ in logarithmic scales. (b) The total displacement $S$ as a function of the event time at $\beta=1$ 
} 
\label{fig:7} 
\end{center}
\end{figure} 
\begin{figure}
\begin{center}
\includegraphics[width=11cm]{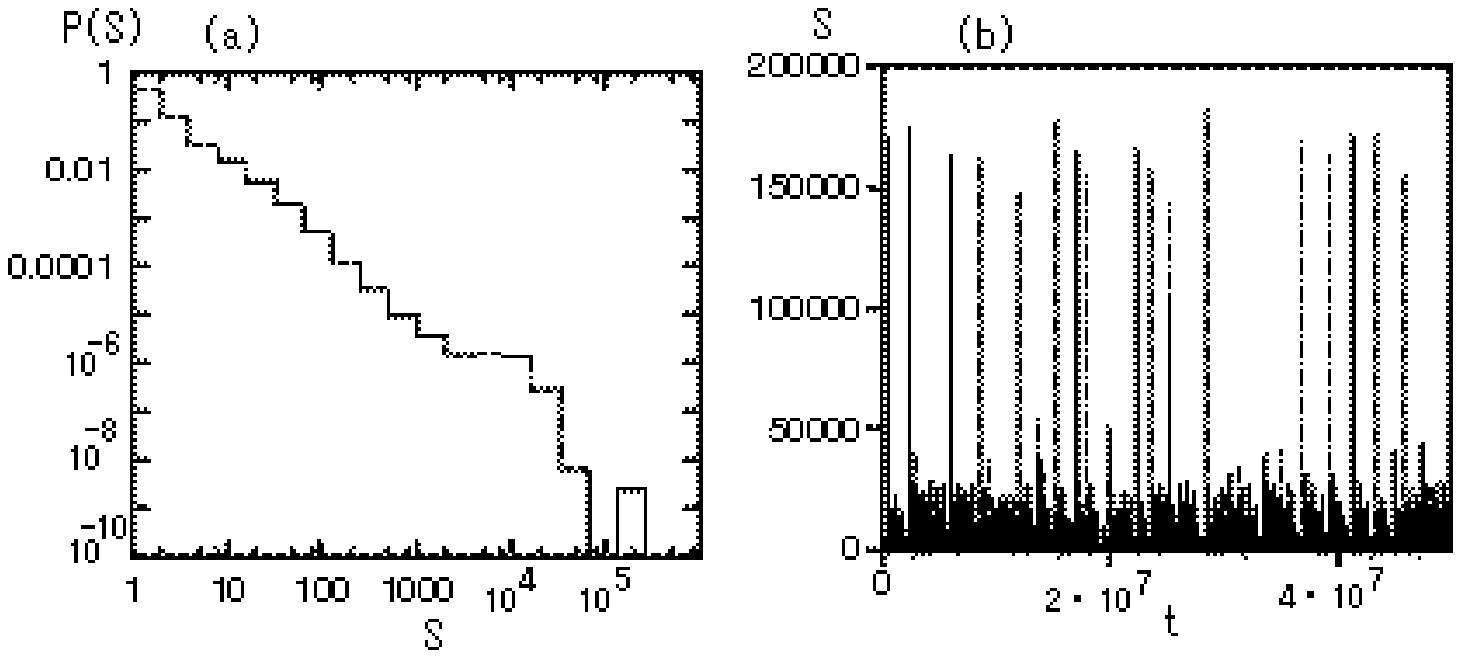}
\caption{(a) Histogram of size distribution of the total displacement $S=\sum_{i=1}^N\Delta x_i$ during each slip event at $\beta=0.2$ in logarithmic scales. (b) The total displacement $S$ as a function of the event time at $\beta=0.2$ 
} 
\label{fig:8} 
\end{center}
\end{figure} 
We have performed numerical simulations with $N=399$ for $K=10,\,D=5,\,V=0.001$ and $F_s=3$.
No-flux boundary conditions are assumed: $x_{N+1}=x_{N},\,\theta_{N+1}=\theta_{N}$ and $\theta_1=\theta_2$.
The initial conditions are $x_i(0)=0, v_i(0)=0.1$ and $\theta_i(0)=0.1$. 
Irregular stick-slip motions are observed initially. A steady sliding state 
is  finally obtained for $\beta>1$ after a long run, however,  stick-slip motions  are 
maintained for $\beta\le 1$.  
Figure 5(a) compares a snapshot pattern of $x_i(t)$ at $t=2\times 10^7$ for $\beta=1$ with the steady sliding solution $x_{i}=Vt-\{1-(1-\beta)V\}/K\cdot \{N^2-(i-N)^2\}/2$. The good agreement implies that the stick-slip state 
is very close to the steady sliding solution at $\beta=1$. 
The average value of the total length  
$L=\langle x_1-x_N\rangle$ is displayed in Fig.~5(b).  For $\beta>1$, the steady 
sliding solution appears and the total length is close to $L=\{1-(1-\beta)V\}N^2/2$. In the stick slip phase, the total length decreases continuously from the value of the steady sliding solution. These results suggest that the transition from the steady sliding phase to the stick slip phase is a continuous transition. Chaotic motion appears at the onset of the instability.  

The slip motion starts at $i=1$ in most cases and the rupture process 
 propagates and ends at a certain  position. 
There are a long interval of stick between two slip motions for small pulling velocity $V$.
The slip events occur in a variety of sizes.  In small slips, the rupture process ends at small $i$. In large slips, the rupture reaches $i=N$. Figure 6 displays time evolutions of a large-size slip near $t=19100$ and an intermediate-size slip near $t=40400$, where the displacement $\Delta x_i=x_i(t+\Delta t)-x_i(t)$ during a time interval $\Delta t=20$ is plotted. The leading edge of the rupture propagates with a nearly constant velocity $c=1.43$ for large $i$, which is fairly smaller than the sound velocity $\sqrt{K}=3.16$.  (Langer and Tang studied rupture propagation in their homogeneous model, epicenters are randomly distributed, and the rupture propagates in both directions from the epicenter.\cite{rf:10} ) 
 
Figure 7 (a) and (b) display a distribution of the event size and the time evolution of the event at $\beta=1$. Here the event size is defined as the total displacement $S=\sum_{i=1}^N\Delta x_i$ during each slip event. For the critical parameter $\beta=1$, the distribution exhibits a power law with exponent nearly 1.5.  Various size of events occur as is seen in Fig. 7(b).  The largest size seems to be determined by the system size.  Figure 7 (a) and (b) display a distribution of the event size and the time evolution of the event at $\beta=0.2$. For $\beta=0.2$, the distribution seems to obey a power law with exponent about 1.6 for small events, however, the distribution deviates from the power law and very large events occur more frequently than is expected from the power law. 
Very large events with similar sizes occur frequently.   
  
\section{Summary and Discussion}
We have proposed a model which exhibits a phase transition from steady sliding to stick slip. For a single block spring system, we have reconfirmed that the time evolution in the slip process hardly depends on the pulling velocity $V$ when $V$ is sufficiently small.  The equation of the time evolution and the period have been evaluated in the limit $V=0$.  Hayakawa evaluated similar quantities for his model through the reduction to the Amontons-Cloulomb model in the limit of $V=0$, neglecting the friction force \cite{rf:5}.  
We have evaluated the period in the limit $V=0$, retaining the friction force.
In a long chain of blocks and springs, the blocks are connected to the neighbors only with harmonic springs, and only the first block is pulled with a constant velocity $V$.  
We have found a continuous transition from steady sliding to stick slip. 
In most cases, slip processes, i.e., ruptures start from the first block. 
At the critical point, the distribution of slip size obeys a power law, however, large events with a typical size occur more frequently than is expected from the power law for small $\beta$, which is consistent with the model of Carlson and Langer.  Our model is a deterministic system, however, it exhibits a dynamical phase transition similar to the thermodynamic phase transition including the critical phenomena.   Bifurcations and chaotic behaviors  were studied in a model of two blocks and the connecting springs  by Vieira \cite{rf:11}, but our model of a long chain of blocks and springs  exhibits a continuous transition from a stationary sliding state to a spatio-temporal chaotic state. 
 Our model suggests that the size distribution depends on a parameter, and the power law distribution is observed near the transition point, although direct relevance to physical systems such as earthquakes is not clear yet.

\end{document}